\documentclass[10pt,twocolumn,letterpaper]{article}
\usepackage[table,xcdraw]{xcolor}
\usepackage[utf8]{inputenc}
\usepackage{algorithm}
\usepackage{algpseudocode}
\usepackage{amsmath} 
\usepackage{cvpr}              

\definecolor{cvprblue}{rgb}{0.21,0.49,0.74}
\usepackage[pagebackref,breaklinks,colorlinks,citecolor=cvprblue]{hyperref}

\def\paperID{*****} 
\def\confName{CVPR}
\def\confYear{2024}

\title{A Closer Look at Spatial-Slice Features Learning for COVID-19 Detection}
\author{$^\dagger$$^1$Chih-Chung Hsu, $^\ddagger$$^1$Chia-Ming Lee, $^2$Yang Fan Chiang,  $^1$Yi-Shiuan Chou, \\$^1$Chih-Yu Jiang,$^1$Shen-Chieh Tai, $^1$Chi-Han Tsai\\
$^1$Institute of Data Science, National Cheng Kung University, Taiwan\\
$^2$Department of Electrical Engineering, National Cheng Kung University, Taiwan\\
{\tt\small $^\dagger$cchsu@gs.ncku.edu.tw, $^\ddagger$zuw408421476@gmail.com}}

\begin{document}
\maketitle
\begin{abstract}

Conventional Computed Tomography (CT) imaging recognition faces two significant challenges: \textbf{(1) There is often considerable variability in the resolution and size of each CT scan}, necessitating strict requirements for the input size and adaptability of models. \textbf{(2) CT-scan contains large number of out-of-distribution (OOD) slices.} The crucial features may only be present in specific spatial regions and slices of the entire CT scan. How can we effectively figure out where these are located? To deal with this, we introduce an enhanced \textbf{S}patial-\textbf{S}lice \textbf{F}eature \textbf{L}earning \textbf{(SSFL++)} framework specifically designed for CT scan. It aims to filter out OOD data within the entire CT scan, enabling us to select crucial spatial slices for analysis by reducing 70\% redundancy totally. Meanwhile, we proposed \textbf{K}ernel-\textbf{D}ensity-based slice \textbf{S}ampling \textbf{(KDS)} method to improve the stability during training and inference stage, therefore speeding up the rate of convergence and boosting performance. As a result, the experiments demonstrate the promising performance of our model using a simple EfficientNet-2D (E2D) model, even with only 1\% of the training data. The efficacy of our approach has been validated on the COVID-19-CT-DB datasets provided by the DEF-AI-MIA workshop, in conjunction with CVPR 2024. Our code is available at \hyperlink{https://github.com/ming053l/E2D}{https://github.com/ming053l/E2D}.

\end{abstract}    
\section{Introduction}
\label{sec:intro}

Computed Tomography (CT) \cite{ct}has become essential in detecting and managing diseases. This technology excels at revealing abnormalities within the body, such as ground-glass opacities and bilateral patchy shadows, which are crucial for the early detection and monitoring of diseases. In diagnosing COVID-19, doctors rely on analyzing lung CT scans of patients. However, since a single patient's CT scan can include hundreds of images, manual examination becomes a time-consuming task, especially when doctors have to evaluate CT scans from dozens or hundreds of patients. This may result in false negatives when dealing with numerous scans.

With the rapid development of deep learning (DL), DL methods \cite{Thyroid,Thyroid2,Thyroid3,Thyroid4,Thyroid5,ml,ml2} have gained prominence for their ability to quickly and accurately identify COVID-19 features while efficiently handling large volumes of data. Furthermore, convolution neural networks (CNNs) have proven to be more effective than methods based on frequency-domain \cite{frequency,frequency2} and low-level features for CT image analysis \cite{sift}.

To address the terribly spreading COVID-19, Kolliaz \emph{et al.} proposed the COVID-19-CT-DB dataset \cite{arsenos2022large,arsenos2023data,kollias2020deep,kollias2020transparent,kollias2022ai,kollias2023deep,kollias2021mia,kollias2023ai}, which encompasses a vast amount of labeled COVID-19 and non-COVID-19 data, advancing the DL methodology and tackling the challenge faced by the huge requirement of high quality dataset for DL-based analysis. Many researchers have designed several methods to deal with COVID-detection task \cite{chen2021adaptive,hsu2022spatiotemporal,hsu2023bag,zhang2021efficient}.

Despite the effectiveness of CT imaging as a tool for detecting abnormalities, it suffers from varying resolutions and quality due to different data servers and scanning machines. The resolution and number of slices in CT images can differ based on the specific scanning machine used, potentially compelling researchers to devise more complex network architectures. Additionally, medical analysis for COVID-19, unlike typical DL-based tasks that focus solely on performance and applications, necessitates maintaining the explainability of model predictions for security and safety reasons \cite{explain1,explain12,chen2021adaptive}.

Inspired by \cite{8578773}, Tran \emph{et al.} presented that factorizing the 3D convolution filters (R3D) into separate spatial and temporal components (R(2+1)D) can yielding significantly gains in accuracy for action recognition. Its effectiveness have been demonstrated by several works on the fields of Video Understanding (VU) \cite{lei2021less,fan2020pyslowfast,lin2019tsm,gberta_2021_ICML} and Human Action Recognition (HAR) \cite{Yang_2020_CVPR,Tran_2019_ICCV}. One video may contains huge redundant information, such as noise from the audio track or each frame, and meaningless background, these factors make it difficult to train the model well \cite{Bhardwaj_2019_CVPR}, resulting in a significant increase in potential costs for data collecting. Likewise, CT scans can be regarded as a special case of video, it also contains various noise resulted from machine aging, and non-important spatial-slice pattern due to its imaging process \cite{ct}. Therefore, the different convolution methods on CT-scan is worthy of discussion.
\begin{figure}
\includegraphics[width=0.48\textwidth]{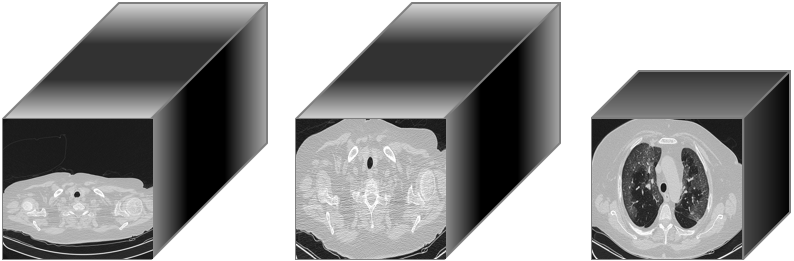}
\caption{The brief illustration for SSFL++. It aim to reduce redundancy in spatial and slice dimension on whole CT-scan to improve model and data quality. (1)Left: original CT-scan. (2)Middle: after reduction at spatial. (3)Right: after reduction at slices.} \label{fig:gradcam.png}
\end{figure}

In this work, we introduce a \textbf{Spatial-Slice Feature Learning (SSFL++)} method, an unsupervised approach designed to reduce computational complexity by effectively removing out-of-distribution (OOD) slices and redundant spatial information. Furthermore, previous works \cite{hsu2023bag,chen2021adaptive} have struggled to identify the most influential slices while considering global sequence information. Based on this observation, we believe there is room for improvement. Therefore, we propose the \textbf{Kernel-Density-based Slice Sampling (KDS)} strategy, which leverages Kernel Density Estimation to simultaneously achieve both objectives. Experimental results have demonstrated our 2D model's outstanding performance, even in the face of data insufficiency.

Our novelties and contributions can be briefly divided into two parts as following mentions: 
\begin{itemize}

\item \textbf{Improved spatial-slice feature learning module:}
SSFL++ is a morphology-based approach for CT scans that removes redundant areas in both spatial and slice dimensions. This significantly reduces computational complexity and efficiently identifies the Regions of Interest (RoI) without the need for complicated designs or configurations. Remarkably, we were able to eliminate 70\% of the area in the COVID-19-CT-DB datasets without any degradation in performance.

\item \textbf{The comparison between 2D, (2+1)D, and 3D for CT-scan is discussed:} To facilitate the development of related research, we conducted a thorough discussion on the use of 2D, 2+1D, and 3D convolutions for CT scan data in COVID-19 detection. Based on experimental results, we believe that the 2D convolutional architecture holds more potential for future applications compared to 3D and 2+1D convolutions.

\item \textbf{Density-aware slice sampling method:}
Coupled with SSFL++'s ability to adaptively remove redundant spatial areas and slices, KDS further adaptively samples the most crucial slices while preserving global sequence information. This approach enhances data efficiency and strengthens the model's few-shot capabilities. Experimental results have shown that our E2D model maintains strong and robust performance under scenarios with few CT scans and slices.

\end{itemize}

\section{Related Work}
\label{sec:relatedwork}

In this section, we introduce the related works on COVID-19 recognition in recent years, along with traditional spatial-temporal feature learning for Video Understanding (VU) and Human Action Recognition (HAR). The philosophy behind these approaches is important for analyzing CT-Scans.

\subsection{Region of Interests for Computed Tomography}
\textbf{Background.} CT \cite{ct} harnesses X-rays, which encircle a specific plane of the human body, while detectors on the opposite side capture the resultant signals. This technique exploits the differential attenuation of X-rays by various tissues, combined with signals obtained from multiple irradiation angles traversing the body, to compile a sinogram. This sinogram facilitates the reconstruction of cross-sectional imagery \cite{BARRETT1984217,Beatty2012TheRT}. Nonetheless, the CT imaging paradigm, necessitating multi-angular signal acquisition for reconstruction, engenders scans replete with extraneous data, potentially escalating labor costs.

Although this technology has been around for a long time, designing a robust and reliable Region of Interest (RoI) selection algorithm for CT scans remains an open problem. Noise and redundancy harm model performance. In recent years, most researchers have still focused on enhancing the feature extraction pipeline \cite{ct1}, or improving the quality of image reconstruction \cite{ct2}, to address the aforementioned challenges. Cobo \emph{et al}. \cite{ct3} suggested that standardizing medical imaging workflows could improve the performance of radiomics and deep learning systems. Jensen \emph{et al}. \cite{ct4} proposed enhancing the stability of CT radiomics with parametric feature maps. Gaidel \emph{et al}. \cite{GAIDEL2017258} introduced a greedy forward selection-based method for lung CT images, but its development was limited due to a lack of robustness against data shifting and noise.

\subsection{COVID-19 Recognition}

In recent years, substantial progress has been achieved in developing methods for COVID-19 recognition. Kollias \emph{et al}. \cite{kollias2020deep} have contributed to this field by analyzing the prediction results of deep learning models based on latent representations. Chen \emph{et al}. \cite{chen2021adaptive} integrated maximum likelihood estimation with the Wilcoxon test, adopting a statistical learning perspective to adaptively select slices and design models with explainability. 

Furthermore, Hou \emph{et al}. proposed a method based on contrastive learning to enhance feature representation. Turnbull \emph{et al}. applied a 3D ResNet \cite{ResNet} for COVID-19 severity classification.  Hsu \emph{et al}. \cite{hsu2022spatiotemporal} introduced a two-step model that combines 2D feature extraction with an LSTM \cite{hochreiter1997long} and Vision Transformer \cite{dosovitskiy2020image}. They presented a 2D and (2+1)D approach \cite{hsu2023bag}, achieving outstanding results in the AI-MIA 2023 COVID-19 detection competition.

\subsection{Spatiotemporal Feature Learning for Video}

Video analysis is crucial for computer vision, as videos contain far more information than single images. This analysis focuses on extracting spatiotemporal features, with traditional methods relying on optical flow \cite{tang2019hallucinating,10.1007/978-3-540-24673-2_3} and trajectory analysis \cite{10.1007/978-3-319-10602-1_38,tra}. With the advent of deep learning (DL), a strategy employing 2D Convolution Neural Networks (CNNs) was proposed \cite{pmlr-v48-fernando16,Ng_2015_CVPR}. This strategy includes temporal feature pooling to aggregate features from different frames for classification. Subsequently, approaches combining CNNs with Recurrent Neural Networks (RNNs) and Long Short-Term Memory (LSTM) networks \cite{lstt} were introduced, aiming to capture the long-range dependencies across various frames. 3D convolution kernels (C3D \cite{Tran_2015_ICCV}, I3D \cite{8099985}) are used in video understanding, capturing channel interactions and local interactions simultaneously. However, they lead to a computational burden and have been regarded as an inefficient approach.

Subsequently, strategies offering greater efficiency were introduced, such as the Non-local network \cite{Wang_2018_CVPR}, S3D \cite{Xie_2018_ECCV}, CoST \cite{li2019collaborative}, SlowFast \cite{feichtenhofer2019slowfast}, and CSN \cite{Tran_2019_ICCV}. These methods more efficiently learn the spatiotemporal features of videos by either reducing the number of sampled frames or replacing the use of 3D convolution with (2+1)D convolution. The prevailing consensus has moved away from the necessity of utilizing a large number of video frames or 3D convolution as the optimal approach for learning spatiotemporal features. Similarly, considering the resemblance between CT scans and videos, it is plausible to learn the feature representation of CT scans using only a small number of slices, without relying on 3D CNNs.

\section{2D, (2+1)D, 3D Convolutions for CT Scan}

In this section, we discuss the three types of convolutions within framework of COVID-19 detection. The detailed architecture is described in Section 5-1.


\textbf{2D: 2D Convolution over the sampled slices.} The use of 2D convolution networks for extracting spatio-temporal features from 3D-cube data faces certain limitations, such as the requirement for strong spectral band or temporal continuity. Without these prerequisites, 2D convolutions may struggle to perform effectively due to their focus on spatial features and a lack of comprehensive sequence modeling. In applications involving CT scans, 2D convolutions are generally considered less effective compared to architectures like 2+1D convolutions, CNN-LSTM, or CNN-RNN, which are capable of capturing spatiotemporal features more efficiently. However, previous 2D CNN approaches often involve pre-processing, where crucial slices are selected and sampled to serve as inputs for the network. This sampling process tends to be overly simplistic, for instance, by manually selecting slices with the least artifacts or best quality, or randomly selecting a few slices to train a 2D CNN model. This limits the network’s potential for global sequential modeling.



\textbf{(2+1)D: 1D Convolution over the extracted features on different dimension.} The 2+1D model is widely regarded as the greatest solution for CT analysis due to its exceptional performance and lower computational costs compared to 3D models. Typically, the 2+1D model performs best as it first extracts features on the spatial scale before modeling the sequences of these extracted features, effectively achieving both. However, according to our experiments, it tends to be less robust in situations with limited samples. This is because CT scans vary greatly in terms of resolution or the number of slices, making the 2+1D model more sensitive to the quantity of training data compared to 2D models. Additionally, we believe a potential concern with the 2+1D model is its difficulty in augmentation since spatial features are encoded into the latent space, the implicit learning approach limits its scalability and interpretability in clinical applications.



\textbf{3D: 3D Convolution over the contiguous slices.} \emph{Compared with 2D and 2+1D, 3D is a heavy computational resource burden for COVID-19 detetion.} The differences between CT scans and conventional videos lie in several key aspects. Firstly, videos typically contain a significantly larger number of frames compared to the number of slices in a CT scan. Secondly, videos enhance their spatio-temporal coherence through frame rates (FPS), whereas the spatial relationships between slices in CT scans are relatively weaker. Lastly, slices in CT scans often exhibit redundancy at the beginning and end, which does not substantially contribute to analysis.

In conclusion, the advantages and weaknesses of these three methods can be itemized as follows:
\begin{itemize}

\item \textbf{2D:}
Training and testing pipeline are simple. The model is robust no matter when few-scan or few-slice. Easy to augment. There are multiple methods which provide an explaniability for 2D model's prediction, such as GradCAM++ \cite{gradcam++}, SHAP \cite{NIPS2017_7062}. Uneasy to capture sequential information unless dedicated design.
\item \textbf{(2+1)D:}
The performance is optimal when there is enough training data and the length of the CT slice sequence is sufficiently large, allowing it to capture sequential information. However, it becomes unstable with only a few scans or slices; the pipeline is complicated. It is also difficult to explain and augment.

\item \textbf{3D:} Training and testing pipeline are simple. Can capture sequential information. Worst performance. Highest computational complexity. Unstable when few-scan and few-slice. Hard to explain and augment.

\end{itemize}

We believe 2D-CNNs have the potential to become mainstream for COVID-19 detection tasks. To enhance the ability of 2D-CNNs to learn sequence information from CT scans, we have designed the KDS method. This approach helps overcome the limitations of 2D-CNNs in this regard, with details to be introduced in Section 4.2.

\section{Methodology}

\label{sec:method}

\subsection{Spatial-Slice Feature Learning}

In this section, we introduces our proposed SSFL++, which aim to figure out the RoIs in spatial and slice dimension, mainly based on the simple but effective computed morphology method and formulation of optimization problem.

\begin{figure}
\includegraphics[width=0.48\textwidth]{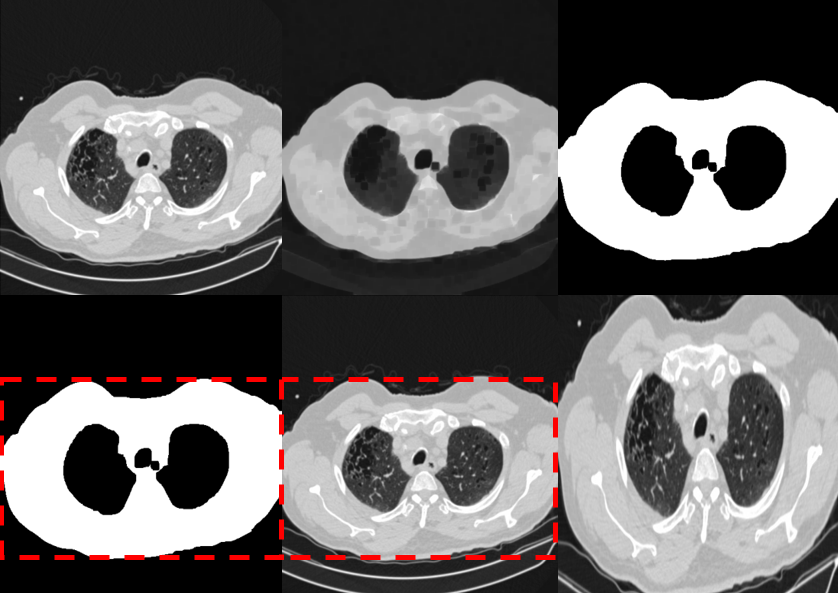}
\caption{The illustration of spatial steps in proposed SSFL++.} \label{fig:spatial.png}
\end{figure}

\textbf{Spatial Steps.} The most importance concern is that CT-scan alway exists large black area between every single CT slice's background, and it will distort the RoI area when resizing to fixed shape to neural network, leading to feature vanish. In order to deal with this, a low-pass filter with a window size of $k\times k$ is applied to all CT slices $\mathbf{Z}$ to eliminate a noises. The low-pass filtering operator can be defined as:

\begin{equation}
\mathbf{Z}_\text{filtered}(i, j) = \frac{\sum_{p=-k}^{k} \sum_{q=-k}^{k} w(p, q) \times \mathbf{Z}(i+p, j+q)}{\sum_{p=-k}^{k} \sum_{q=-k}^{k} w(p, q)}
\end{equation}
where $w(p,q)$ represents the weight at position $(p, q)$ in the filter kernel. The above formula can determine the segmentation $\mathbf{Mask}$ of the filtered slices by a threshold $t$:

\begin{equation}
    \label{eq:seg}
    \mathbf{Mask}[i,j] = 
    \begin{cases}
    0,\,\text{if}\,\mathbf{Z}_\text{filter}[i,j] < t\\
    1,\,\text{if}\,\mathbf{Z}_\text{filter}[i,j] >= t
    \end{cases}
\end{equation}
where i, j denote as an pixel for every single CT slice $\mathbf{Z}^{c}$, which resolution is $x$ $\times$ $y$. A Cropped region $\mathbf{Z}_\text{crop}^{c}$ can be calculated by:

\begin{equation}
\begin{aligned}
    \text{min}(\mathbf{Z}_\text{crop}^{c}(x)) = \min\{i \mid \mathbf{Mask}[i, j] = 1, \forall i\}\\
    \text{max}(\mathbf{Z}_\text{crop}^{c}(x))= \max\{i \mid \mathbf{Mask}[i, j] = 1, \forall i\}\\
    \text{min}(\mathbf{Z}_\text{crop}^{c}(y)) = \min\{j \mid \mathbf{Mask}[i, j] = 1, \forall j\}\\
    \text{max}(\mathbf{Z}_\text{crop}^{c}(y)) = \max\{j \mid \mathbf{Mask}[i, j] = 1, \forall j\}
    \end{aligned}
\end{equation}

$\mathbf{Z}_{crop}^{c}$ is yielded accordingly, we can further resize the resolution of $\mathbf{Z}_\text{crop}^{c}$ to $H$$\times$$W$ for the slice steps and as an input of neural network. Spatial Steps in proposed 4SFL effectively filter out non-lung tissue regions (also known as RoIs in spatial dimension), and reduce computational complexity, as the Figure \ref{fig:spatial.png} illustrated.

\begin{figure}
\includegraphics[width=0.48\textwidth]{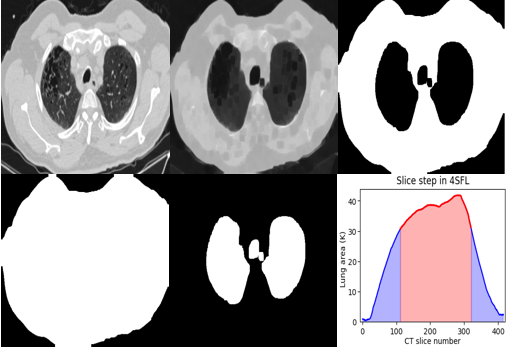}
\caption{The illustration of slice steps in proposed SSFL++. The line graph in the bottom right corner represents the area of each slice in a CT scan. The blue area denotes OOD data that have been removed, while the red area represents the CT slices that have been selected.} \label{fig:slice.png}
\end{figure}

\textbf{Slice Steps.} To find the lung tissue region in the CT scan, we used the binary dilation algorithm \cite{enwiki:1082436538} to obtain the filled result $\mathbf{Mask}_\text{filled}$. The difference between the $\mathbf{Mask}$ and filled mask $\mathbf{Mask}_\text{filled}$ represents the lung tissue region. The above method can be summarized as the following formula:

 \begin{equation}
    \label{area}
    Area(\mathbf{Z}) = \sum_i\sum_j\mathbf{Mask}_\text{filled}(i,j) - \mathbf{Mask}(i,j).
\end{equation}

After the above technique, we can finally obtain a range where $s$ and $e$ denote the starting and ending indexes, respectively, and $n_c$ is the constraint of the number of slices for a single CT scan to select most importance RoIs in slice dimension with proportion $\alpha$. The optimization problem can be formulated as following:

\begin{equation}
    \label{eq:area}
    \begin{split}
    & \underset{s,\,e}{\text{maximize}} \quad \sum^e_{i=s}Area(\mathbf{Z}_i), \\
    & \text{subject to } \quad e-s \leq n_c, \\
    & \quad \frac{\sum^e_{i=s}Area(\mathbf{Z}_i)}{\sum_{i=1}^{n_c} Area(\mathbf{Z}_i)} \geq \alpha.
    \end{split}
\end{equation}

It is worth noting that we sort all CT slices according to their slice numbers $n_c$, as illustrated in the bottom-right corner of Figure \ref{fig:slice.png}.

The spatial and slice steps of proposed SSFL++ follow unsupervised learning manner, which only follow the prior knowledge of lung-CT-scan. It can be generalize to other organs or body parts CT-scan. However, it may require parameter adjustments based on their specific characteristics.
Additionally, with the SSFL++, the visual explanation method can also look RoI more concentrated, as shown in Figure \ref{fig:gradcam.png}.

\begin{figure}
\includegraphics[width=0.48\textwidth]{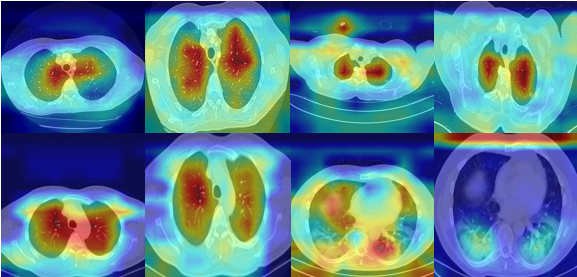}
\caption{The GradCAM++ \cite{gradcam++} visualization before and after proposed SSFL++. By reducing redundancy on the spatial scale, we can implicitly enhance the visual effectiveness of Explainable AI, thereby facilitating clinical applications.} \label{fig:gradcam.png}
\end{figure}

\subsection{Density-aware Slice Sampling}

\begin{figure*}
\includegraphics[width=1\textwidth]{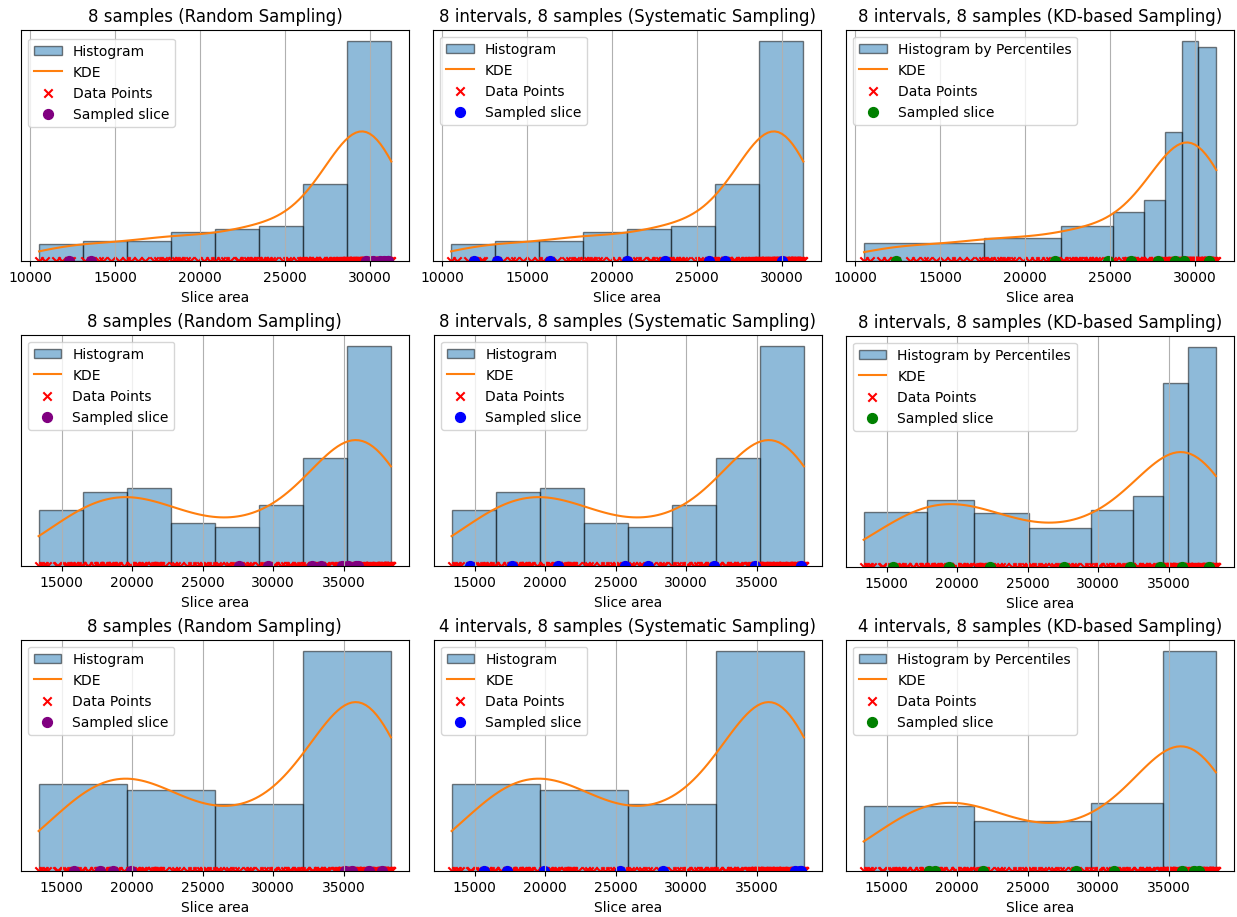}
\caption{The comparison between random sampling, systematic sampling, and the proposed KDS method is noteworthy. As illustrated, random sampling fails to uniformly sample CT slices of varying area sizes, tending to select larger areas while neglecting global information. This results in greater bias and randomness during training and inference. On the other hand, systematic sampling divides the area into equally lengthened sub-intervals before randomly selecting samples from them. Although this approach can capture global information, it is ineffective at sampling the most crucial CT slices. Our proposed KDS method combines the advantages of both methods without their drawbacks, achieving a better balance. KDS can implicitly improve data efficiency, thereby enhancing the model's few-shot capability.} \label{fig:kde.png}
\end{figure*}

\textbf{Background.} The SSFL proposed by Hsu \emph{et al.} \cite{hsu2023bag} employs a random sampling method to select slices, which were used for the detection of COVID-19 using 2D and 2+1D CNNs. However, random sampling may potentially introduce bias and instability when training and inference, and it does not efficiently identify the most representative CT slices, as shown in Figure \ref{fig:kde.png}. 

In order to address this, we propose a Kernel-Density-based Slice Sampling (KDS). It performs kernel density estimation (KDE) on the selected slices-set [$\mathbf{Z}_e$,$\mathbf{Z}_s$], adaptively and wisely sampling the most crucial CT-slices. Meanwhile, it also keeps the sequence information globally and alleviates the instability during training and inference stage.

\textbf{Definition.} KDE is a classic method to estimate the probability density function (PDF) of a random variable in a non-parametric manner. It can be defined as:

\begin{equation}
    {\widehat {f}}_{h}(x)={\frac {1}{s}}\sum _{i=1}^{s}K_{h}(x-x_{i})={\frac {1}{sh}}\sum _{i=1}^{s}K\left({\frac {x-x_{i}}{h}}\right)
\end{equation}

\begin{equation}
K(x, x') = \exp\left(-\frac{\|x - x'\|^2}{2\sigma^2}\right),
\end{equation}
where $h$ is the bandwidth constant, calculated by Scott-rule \cite{Scott}, $K$ is a Gaussian kernel, $s$ is a smooth factor of estimated density function, (the higher the smoother, we set it to 100). For a given KDE, we can create several sub-intervals by calculating its Cumulative Distribution Function (CDF), where the length of each sub-interval adaptively changes with its $p$-percentile. The CDF of KDE and its $p$-percentile can be calculated as following formulas:

\begin{equation}
F(x) = \int_{-\infty}^{x} \hat{f}_{h}(t) dt, 
F(q_p) = p
\end{equation}

\begin{figure}
\includegraphics[width=0.45\textwidth]{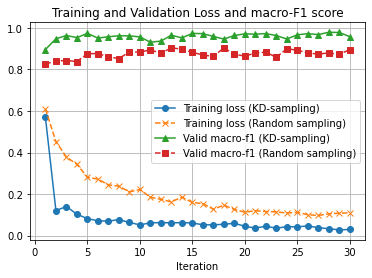}
\caption{In terms of optimizing procedure, our proposed KDS approach, compared to the random sampling used by Hsu \emph{et al.} \cite{hsu2023bag}, is more capable of learning the global information of CT scans, thereby accelerating the convergence rate and enhancing the model performance.} \label{fig:kdsloss.png}
\end{figure}

In the proposed KDS method, we determine the probability of slices being selected in each interval based on the density from KDE, while also ensuring that each sub-interval has at least one sample selected. This method captures the global sequential information and increases the probability of selecting the most crucial CT slices.
\section{Experiment}
\label{sec:experiment}

\begin{table*}[]
    \centering
    \scalebox{0.925}{
    \begin{tabular}{l|ccc|ccc|ccc}
    \multicolumn{1}{c|}{} & \multicolumn{3}{c|}{Spatial Area (K)} & \multicolumn{3}{c|}{Slice Length} & \multicolumn{2}{c|}{Spatial $\times$ Slice (M)} &  Total \\\cline{2-10}
    \multicolumn{1}{c|}{} & Before & After & $\Delta$ $(\%)$ & Before & After & $\Delta$ $(\%)$ & Before & \multicolumn{1}{c|}{After} & $\Delta$ $(\%)$ \\ \hline
    Training & 267.25 & 155.53 & 0.4184 & 285.32 & 142.91 & 0.4983 & 76.25 & \multicolumn{1}{c|}{22.22} & 0.7085 \\
    Positive & 266.42 & 157.69 & 0.4088 & 295.90 & 148.18 & 0.4985 & 78.83 & \multicolumn{1}{c|}{23.36} & 0.7036 \\
    Negative & 268.21 & 153.03 & 0.4296 & 273.97 & 137.26 & 0.4981 & 73.48 & \multicolumn{1}{c|}{21.00} & 0.7141 \\ \hline
    Validation & 265.62 & 155.23 & 0.4172 & 281.95 & 141.23 & 0.4984 & 74.89 & \multicolumn{1}{c|}{21.92} & 0.7072 \\
    Positive & 268.94 & 160.48 & 0.4061 & 280.53 & 140.55 & 0.4984 & 75.45 & \multicolumn{1}{c|}{22.55} & 0.7010 \\
    Negative & 262.12 & 149.69 & 0.4288 & 283.49 & 141.97 & 0.4984 & 74.30 & \multicolumn{1}{c|}{21.25} & 0.7139 \\ \hline
    (T+V) Positive & 267.25 & 155.53 & 0.4184 & 292.96 & 146.72 & 0.4985 & 78.29 & \multicolumn{1}{c|}{22.81} & 0.7085 \\
    (T+V) Negative & 267.01 & 152.37 & 0.4294 & 275.78 & 138.16 & 0.4982 & 73.64 & \multicolumn{1}{c|}{21.05} & 0.7141 \\ \hline
    Total & 266.94 & 155.47 & 0.4182 & 284.68 & 142.59 & 0.4983 & 75.99 & \multicolumn{1}{c|}{22.16} & 0.7082 \\ \hline
    Testing & 279.55 & 153.41 & 0.4520 & 309.39 & 154.67 & 0.5003 & 86.48 & \multicolumn{1}{c|}{23.72} & 0.7256 
    \end{tabular}
}
\caption{The reduction in redundant data achieved by the SSFL++ module is evaluated across three dimensions: spatial, slice, and overall. This approach quantifies the efficiency of the SSFL++ module in reducing unnecessary information in CT scans, enabling more focused analysis and processing. By minimizing data redundancy, the module enhances computational efficiency and potentially improves the accuracy of subsequent analyses or models applied to the CT data.}
\label{tab:4sfl}
\end{table*}

\hspace{\parindent} \textbf{Dataset description.} In our experiments, we used a total of 1,684 COVID-19-CT-DB data, provided by Kollias \emph{et al.} \cite{cvpr24}. The dataset information have shown in Table \ref{tab:dataset_subgroup1}. Our loss function is binary cross-entropy. In order to ensure stability and fairly check performance during the experiments, group-5-fold-cross-validation is used. Data augmentation and hyperparameters are kept consistent in all experiments.

\begin{table}[!ht]
    \centering
    \begin{tabular}{c | c c |c} 
       Type& Positive Scan& Negative Scan&   Total Scan \\\hline

      Training     &       703&         655&     1358\\ 

      Valid     &       170&          156&      326\\\hline
      Total     &       873&          811&      1684\\\hline
      Testing     &       -&          -&      1413\\\hline

      Type & Positive Slice & Negative Slice &   Total Slice\\\hline
            Training     &       206608&         178722&     385330\\
      Valid     &       46042&          43679&      89721\\\hline
      Total     &       252650&          222401&      475051\\\hline
      Testing     &       -&          -&      437185\\
    \end{tabular}
    \caption{The number of data samples at the scan and slice level.}
    \label{tab:dataset_subgroup1}
\end{table}

\textbf{Hyperparameter settings.} The Adam \cite{Adam} optimizer was used with a learning rate of $1e-4$ and a weight decay of $5e-4$. The batch-size is set to $16$.

\textbf{Data Augmentation.} In our experiments, we utilized common augmentation strategy like HorizontalFlip, RandomScaleShifting to prevent overfitting and enlarge feature space. Additionally, we find that HueSaturationValue, RandomBrightnessContrast and CoarseDropout \cite{devries2017improved} are also used.

\textbf{Evaluation Metric.} We mainly used F1-score in the experiments for model evaluation. F1-score is a metric used to determine the accuracy of a binary classification model. It combines the harmonic mean of Precision and Recall.

\begin{equation}
\text{f1-score} = 2 \times \frac{\text{precision} \times \text{recall}}{\text{precision} + \text{recall}},
\end{equation}
where precision and recall are computed for COVID and non-COVID. The macro f1-score is the average of the f1-scores for all classes:
\begin{equation}
\text{macro f1-score} = \frac{1}{N} \sum_{i=1}^{N} \text{f1-score}_{i}
\end{equation}
where $N$ is the number of classes, and $\text{f1-score}_i$ is the f1-score for the $i$-th class. These metrics provide a balanced evaluation of the model's ability to classify each class accurately and its overall performance across all classes.

\subsection{Model Details and Performance Comparison} To provide a more comprehensive comparison and improve future research, we designed simple E2D, E2+1D, E3D in our experiments. The backbones are all based on \textbf{E}fficientNet-b3 \cite{efficientnet,rw2019timm}. The baseline method and detailed pipeline are as follows:

\textbf{Baseline}: The baseline method is presented in \cite{cvpr24}, Kollias \emph{et al.} adopted CNN-RNN to extract feature within all CT-slice. First, all CT-slices are resized to $224$ $\times$ $224$ to extract feature, then RNN (GRU \cite{chung2014empirical} with $128$ neurons) analyzed the 2D-CNN (ResNet-50 \cite{ResNet}) features. The output of the RNN element is then forwarded to a fully connected layer. In addition, this also includes a dropout layer (the dropout rate is set to $0.8$) before the fully connected layer.

\textbf{E2D}: From the CT-scans processed by SSFL++, subsequently, we use our proposed KDS. These sampled slices are resized to $384$ $\times$ $384$ and extracted to high-representation features.

\textbf{E2+1D}: Similar to E2D, firstly, the CT scans processed by SSFL++ are resized to $384$ $\times$ $384$. And $100$ slices are selected to be encoded. therefore, we used 2D encoder to get an encoded vectors. By doing so, the CT scans will be encoded into latent feature queue, which size is $224$ $\times$ $100$. Subsequently, we randomly sampled $50$ features from latent feature queue, and utilized a simple 1D convolution with kernel size $1$ $\times$ $1$ in $e$ or $l$ dimensions to capture sequential information.

\textbf{E3D}: We first utilized SSFL++ to remove OOD slices and redundant spatial information, and then sample a certain number of CT slices for modeling. 

The experimental results, as presented in Table \ref{tab:fewshot}, highlight the E2D model's exceptional performance when paired with KDS on the COVID-19 database 2024 validation set. It also showcases remarkable robustness in few-scan scenarios, delivering results that instill confidence. Comparatively, the E2D model utilizing KDS achieves a significant improvement in scan-level f1-score compared to its counterpart that employs random sampling. This underscores the capability of 2D convolutions to implicitly capture global sequence information through an appropriate sampling method. In contrast, the E3D model demands a large sample size, resulting in limited performance and higher computational requirements.

\begin{table}[]
\scalebox{0.8}{
\begin{tabular}{c|cc|c|c}
Model type & Scans& Sampled slice & \begin{tabular}[c]{@{}c@{}}macro f1-score\\ (slice-level)\end{tabular} & \begin{tabular}[c]{@{}c@{}}f1-score\\ (scan-level)\end{tabular} \\ \hline
\begin{tabular}[c]{@{}c@{}}baseline \cite{cvpr24}\end{tabular} & 100\% & - & - & 78.00 \\ \hline
E3D & 1\% & 33(random) & - & 32.55 \\
 & 50\% & 33(random) & - & 78.54 \\
 & 100\% & 33(random) & - & 86.76 \\
 & 100\% & 50(random) & - & 87.05 \\ \
 & 100\% & 80(random) & - & 90.24 \\
 & 100\% & 120(random) & - & 91.05 \\ \hline
E(2+1)D & 1\% & 8(random) & 73.46 & - \\
 & 50\% & 8(random) & 87.64 & - \\
 & 100\% & 8(random) & 91.39 & - \\
 & 100\% & 16(random) & 92.31 & 93.69 \\ \hline
E2D & 1\% & 8(random) & 88.94 & 92.11 \\
 & 50\% & 8(random) & 91.52 & 92.42 \\
 & 100\% & 8(random) & 92.44 & 93.18 \\
 & 100\% & 16(random) & 92.68 & 93.37 \\ 
\multicolumn{1}{l|}{} & 1\% & 4(KDS) & 91.42 & 96.42 \\
\multicolumn{1}{l|}{} & 1\% & 8(KDS) & 91.88 & 99.80 \\
\multicolumn{1}{l|}{} & 100\% & 8(KDS) & 93.46 & \textbf{100.00} \\
\multicolumn{1}{l|}{} & 100\% & 16(KDS) & \textbf{94.11} & \textbf{100.00}
\end{tabular}
}
\caption{Performance comparison between baseline provided by Kollias \emph{et al.} \cite{cvpr24}, and proposed E2D, E2+1D, E3D on COVID-19-CT-DB validation set.}
\label{tab:fewshot}
\end{table}

\subsection{Ablation Study}

\begin{table}[]
\scalebox{0.85}{
\begin{tabular}{ccc|c|c}
Spatial step & Slice step & KDS & \begin{tabular}[c]{@{}c@{}}marco f1-score\\ (slice level)\end{tabular} & \begin{tabular}[c]{@{}c@{}}f1-score\\ (scan level)\end{tabular} \\ \hline
\multicolumn{1}{l}{} & \multicolumn{1}{l}{} & \multicolumn{1}{l|}{} & 80.41 & 81.26 \\
\checkmark &  &  & 88.01 & 88.04 \\
 & \checkmark &  & 90.32 & 90.48 \\
\checkmark & \checkmark &  & 92.68 & 93.37 \\ \hline
\checkmark & \checkmark & \checkmark & \textbf{94.11} & \textbf{100.00}
\end{tabular}}
\caption{The ablation study of proposed SSFL++ and KDS on COVID-19-CT-DB validation set.}
\label{tab:ablation}
\end{table}

\begin{table}[]
\scalebox{0.85}{
\begin{tabular}{l|c|c|c}
 & macro-F1 & F1(NONCOVID) & F1(COVID) \\ \hline
baseline \cite{cvpr24} & 85.11 & 87.48 & 82.74 \\ \hline
\textbf{E2D (Ours)} \cite{hsu2024simple} & \textbf{94.39} & \textbf{95.52} & \textbf{93.26}
\end{tabular}}
\caption{The results on COVID-19-CT-DB testing set.}
\label{tab:my-table}
\end{table}
To further analyze the impact of SSFL++ and KDS on the COVID-19 detection task, the ablation study were conducted, with results presented in Table \ref{tab:ablation}. All experiments are based on the E2D model, with all experimental hyperparameters kept constant. The results demonstrate that the proposed SSFL++ significantly enhances performance, implying the importance of spatial redundancy in CT scans and efficient slice selection. On the other hand, KDS further improves the model's prediction ability at the slice-level and makes significant progress at the scan-level, achieving convincing performance. KDS effectively addresses the lack of global sequential modeling capability in 2D-CNN when analyzing CT images.

\section{Generalizability}

\label{sec:generalizability}

\begin{figure}
\center
\includegraphics[width=0.45\textwidth]{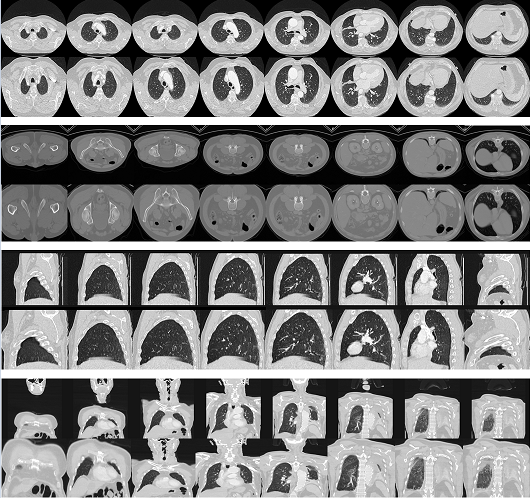}
\caption{CT slices from different views and body parts, as well as the results after processing through the spatial step in our proposed SSFL++, are presented. From left to right, the sequence represents the process of CT imaging, where OOD data tend to concentrate at the beginning and the end. The middle section represents the RoI area. As shown in the figure, SSFL++ performs well under various conditions.} \label{fig:ge.png}
\end{figure}

Our proposed SSFL++ not only excels in performance on the  COVID-19-CT-DB \cite{cvpr24} but also demonstrates commendable efficacy on CT scans from various views and body parts. We showcased the versatility of SSFL++ by selecting four distinct types of data, with the results depicted in Figure \ref{fig:ge.png}. From top to bottom, the figures represent the different views or body parts before and after SSFL++. Specifically, (a) (c) (d) are lung CT scans from the COVID-19-CT-DB dataset, featuring the axial, sagittal, and coronal views. Meanwhile, (b) involves a dataset provided by \cite{aocr2024}, aimed at identifying acute appendicitis from CT scans of acute abdomen cases.

Additionally, it is important that when using SSFL++ on CT slices of different body parts or from different views, its hyperparameters may need specific adjustments. For instance, in the case of (b), the original settings might select OOD slices rather than the RoI slices.

\section{Conclusion}

\label{sec:conclusion}

%

We conducted a comprehensive analysis of the COVID-19 detection task, noting that CT scans often contain a large amount of redundant information, which limits the performance of models. To address this issue, we introduced a simple morphology-based method for CT images, named Spatial-Slice Feature Learning (SSFL++), designed to efficiently and adaptively locate the Region of Interest (RoI). This method effectively reduces redundancy across both spatial and slice dimensions. Furthermore, to inspire future research, we analyzed the advantages and disadvantages of 2D, 2+1D, and 3D convolutions on CT data. After extensive experimentation, we believe that 2D-CNNs hold the greatest potential in the wild.

To overcome the limitations previously encountered by 2D-CNN in research, we combined SSFL++ with the further designed KDS, thereby addressing the instability brought about by random sampling during the training and inference. Moreover, through the global sequence modeling, we activated the potential of 2D-CNNs. Finally, our method demonstrated promising results on the validation and testing sets provided by the DEF-AI-MIA workshop.

{
    \small
    \bibliographystyle{ieeenat_fullname}
    \bibliography{main}
}


\end{document}